\documentclass[pdftex,twocolumn,epjc3]{svjour3}          

\RequirePackage[T1]{fontenc}
\smartqed  

\RequirePackage{graphicx}
\RequirePackage{mathptmx}      
\RequirePackage{flushend}
\RequirePackage[numbers,sort&compress]{natbib}
\RequirePackage[colorlinks,citecolor=blue,urlcolor=blue,linkcolor=blue]{hyperref}

\usepackage{subfiles}
\usepackage{tablefootnote}
\usepackage[switch]{lineno}
\usepackage{enumitem}
\usepackage{siunitx} 
\usepackage{amsmath}
\usepackage{orcidlink} 
\setlist[itemize]{label={$\bullet$}} 
\usepackage{hyphenat}
\usepackage{gensymb} 
\usepackage{multirow} 
\usepackage{array}
\usepackage{tabularx}

\providecommand{\pcb}{P_{corr,B}}
\providecommand{\pf}{P_{\rightarrow}}
\providecommand{\pb}{P_{\leftarrow}}

\providecommand{\pix}{P_{DiCT}}

\providecommand{\drr}{\widetilde{R_R}} 
\providecommand{\nxe}{N_{Xe}}

\journalname{Eur. Phys. J. C}

\begin{document}

\title{Characterization of external cross-talk from silicon photomultipliers in a liquid xenon detector}


\author{
D.~Gallacher~\thanksref{addr1,e1}~\orcidlink{0000-0002-9395-0560}  \and
A.~de~St.~Croix~\thanksref{addr2, addr6,e2}~\orcidlink{0000-0002-6517-420X} \and
S.~Bron~\thanksref{addr2} \and
B.~M.~Rebeiro~\thanksref{addr1}~\orcidlink{0000-0003-1481-5385} \and
T.~McElroy~\thanksref{addr1} \and
S.~Al~Kharusi \thanksref{addr1,i1} \and
T.~Brunner~\thanksref{addr1} \and
C.~Chambers~\thanksref{addr1,addr2} \and
B.~Chana~\thanksref{addr4} \and
Z.~Charlesworth~\thanksref{addr2}~\orcidlink{0009-0008-4444-2122}\and
E.~Egan~\thanksref{addr1} \and
M.~Francesconi~\thanksref{addr5} \and
L.~Galli~\thanksref{addr5} \and
P.~Giampa~\thanksref{addr2} ~\orcidlink{0000-0002-3867-2799} \and
D.~Goeldi~\thanksref{addr4} \and
S.~Lavoie~\thanksref{addr1} \and
J.~Lefebvre~\thanksref{addr3} \and
X.~Li~\thanksref{addr2, addr8} \orcidlink{0009-0006-0322-3017} \and
C.~Malbrunot~\thanksref{addr2, addr1,addr7} \orcidlink{0000-0001-6193-6601} \and
P.~Margetak~\thanksref{addr2} \and
N.~Massacret~\thanksref{addr2}~\orcidlink{0000-0002-5694-7661} \and
S.~C.~Nowicki~\thanksref{addr1}~\orcidlink{0000-0003-2497-8057} \and
H.~Rasiwala~\thanksref{addr1} \and
K.~Raymond~\thanksref{addr2} \and
F.~Retière~\thanksref{addr2} \and
S.~Rottoo~\thanksref{addr1}~\orcidlink{0009-0001-2482-4544} \and
L.~Rudolph~\thanksref{addr1} \and
M.~A.~Tétrault~\thanksref{addr3} \and
S.~Viel~\thanksref{addr4}~\orcidlink{0000-0001-9554-4059} \and
N.~V.~H.~Viet~\thanksref{addr3,i3} \and
L.~Xie~\thanksref{addr2}
}

\thankstext{e1}{e-mail: david.gallacher@mail.mcgill.ca}
\thankstext{e2}{e-mail: 20agds@queensu.ca}
\thankstext{i1}{Current location: Stanford University, CA 94305, United States}
\thankstext{i3}{Current location: Research Center for Nuclear Physics, Osaka University, Osaka, Japan}

\institute{Physics Department, McGill University, Montréal, QC, H3A 2T8, Canada \label{addr1}
\and
TRIUMF, Vancouver, BC, V6T 2A3, Canada \label{addr2}
\and
Department of Physics, Carleton University, Ottawa, ON, K1S 5B6, Canada \label{addr4}
\and
Université de Sherbrooke, Sherbrooke, QC, J1K 2R1, Canada \label{addr3}
\and
Istituto Nazionale di Fisica Nucleare Pisa, 56127 Pisa PI, Italy \label{addr5}
\and
Department of Physics, Queen's University, Kingston, ON, K7L 3N6, Canada \label{addr6}
\and
Department of Physics and Astronomy, University of British Columbia, Vancouver, BC, V6T 1Z1, Canada
\label{addr7}
\and
Department of Physics, Simon Fraser University, Burnaby, BC, V5A 1S6, Canada
\label{addr8}
}


\date{Received: date / Accepted: date}

\maketitle

\begin{abstract}
    The Light-only Liquid Xenon experiment (LoLX) employs a small-scale detector equipped with 96 Hamamatsu VUV4 silicon photomultipliers (SiPMs) submerged in 5 kg of liquid xenon (LXe) to perform characterization measurements of light production, transport and detection in xenon. In this work, we perform a novel measurement of the ``external cross-talk'' (ExCT) of SiPMs, where photons produced in the avalanche process escape the device and produce correlated signals on other SiPMs. SiPMs are the photodetector technology of choice for next generation rare-event search experiments; understanding the sources and effects of correlated noise in SiPMs is critical for producing accurate estimates of detector performance and sensitivity projections. We measure the probability to observe ExCT through timing correlation of detected photons in low-light conditions within LoLX. Measurements of SiPM ExCT are highly detector dependent; thus the ExCT process is simulated and modelled using the GEANT4 framework. Using the simulation, we determine the average transport and detection efficiency for ExCT photons within LoLX, a necessary input to extract the expected ExCT probability from the data. For an applied overvoltage of 4~V and 5~V, we measure a mean number of photons emitted into the LXe per avalanche of $0.5^{+0.3}_{-0.2}$ and $0.6^{+0.3}_{-0.2}$, respectively. Using an optical model to describe photon transmission through the SiPM surface, this corresponds to an estimated photon yield inside the bulk silicon of $20^{+11}_{-9}$ and $25^{+12}_{-9}$ photons per avalanche. The relative increase in intensity of SiPM ExCT emission between 4~V and 5~V is consistent with expectation for the linear increase of gain with respect to overvoltage. 
\end{abstract}

\section{Silicon photomultipliers for liquid xenon scintillation light detection}

Development of liquid xenon (LXe) detector technology has grown substantially in recent years, with applications in neutrinoless double beta decay searches \cite{exo200_2019,Adhikari_2021}, direct dark matter search experiments \cite{pandaX_2024,Akerib_2020,xenoncollaboration2024xenonnt}, gamma-ray observatories \cite{2000AIPC.510.799A}, medical applications in nuclear molecular imaging  \cite{ROMOLUQUE2020162397,GallegoManzano2018} and high intensity rare decay searches \cite{pioneercollaboration2022testing,meg2upgrade}. LXe is a scintillating material: energy deposition from incident radiation produces ionization electrons and vacuum ultraviolet (VUV) light with a mean wavelength of 175 nm \cite{FUJII2015293}. Detection of the VUV scintillation light has traditionally been carried out by VUV sensitive photomultiplier tubes (PMTs) or avalanche photodiodes \cite{exo200_2019,ZHENG2020101145}. Future LXe detectors, such as the nEXO experiment \cite{Adhikari_2021}, will employ silicon photomultipliers (SiPMs) \cite{Gallina_2022} as their light detection technology.

\par 
SiPMs are single-photon sensitive light sensors, used in a variety of applications \cite{gundacker2020}, with significant advantages over the traditional choice of PMTs in certain applications due to their low intrinsic  concentration of uranium and thorium, lower operating voltage, insensitivity to magnetic fields, fair detection efficiency in VUV, and excellent timing resolution \cite{Gallina_2022,DELGUERRA2010223}. SiPMs consist of a large array of single-photon avalanche diodes (SPADs) each operated in the Geiger regime. An incoming photon may trigger a charge avalanche in a SPAD, producing a detectable characteristic pulse before being quenched and then reset.

The avalanche process begins when a photon is absorbed in the sensitive region of the SiPM. During the avalanche, observations show that near infra-red light (NIR) is produced \cite{s21175947,raymond2024stimulated}. This NIR light is the source of the correlated noise known as `optical cross-talk' (OCT), and can transmit to neighbouring SPADs in the same device and trigger additional avalanches, referred to as direct OCT or `DiCT'. NIR light may also produce charge carriers outside the depletion region of the device. These charge carriers can then drift into the depletion region and produce delayed avalanches, known as delayed OCT or `DeCT'. Other sources of correlated and uncorrelated noise exist for SiPMs as well, such as dark current (DC), and after-pulsing  (AP) \cite{ACERBI201916,gundacker2020}. Understanding SiPM correlated noise is crucial for experiments that require single photo-electron (SPE) counting resolution, where SPE-equivalent noise from correlated avalanches can have an outsized impact on performance. Fluctuations in correlated noise directly impact energy resolution for scintillation detectors \cite{Gallina_2022}, and must be studied in detail to produce accurate estimates of detector sensitivity.

\par 
In addition to transmitting into neighbouring SPADs, the NIR cross-talk photon can leave the SiPM entirely and travel across the detector to produce a correlated avalanche on a distant SiPM. This type of correlated noise is referred to as external OCT or `ExCT'. The impact of SiPM ExCT on future detectors employing SiPMs is currently under investigation \cite{LEE2024169101,Boulay2023,Gibbons:2023iux,Razeto2024,guan2023study}. Measurements of SiPM ExCT are highly detector dependent,  influenced by factors such as  optics, geometry and SiPM operational voltage, highlighting the need for simulations and effective models to support accurate estimates of intrinsic photosensor response and sensitivity.

In this work, using timing correlations between SiPMs signals we directly measure ExCT for devices immersed in LXe using the Light-only Liquid Xenon experiment (LoLX). We developed a physics model of SiPM ExCT which was used in the LoLX Geant4 simulation to support the  ExCT probability measurement. We developed an effective model parametrizing the observed ExCT signal in terms of simulation informed transport efficiencies and other detector specific parameters. This model is then used to convert measured coincidences within LoLX to ExCT photon yields; the mean number of photons emitted from the SiPM surface per avalanche and the mean number of photons produced within the bulk silicon. These photon yields are especially relevant for other experiments, as they are independent of detector geometry and optics. The effective model and the Geant4 simulation strategy may be applied to other detector simulations to estimate the impact of SiPM ExCT on performance and measurement sensitivity. In addition to the Geant4 ExCT model, we have developed a detailed optical model of the SiPMs used in LoLX to provide accurate estimates of SiPM performance across a broad range of wavelengths and incident angles. LoLX is described in Section~\ref{sec:1}. We present the measurement, effective model and simulations in detail in Section~\ref{sec:2}. The results of this work and sources of systematic uncertainties are given in Section~\ref{sec:3}, and we conclude in Section~\ref{sec:4} with a discussion of our results, the dominant sources of uncertainty and a comparison with other measurements.

\section{Light-only-Liquid Xenon experiment}
\label{sec:1}
LoLX is a small scale LXe detector designed to study the emission, transport and detection of LXe scintillation light using SiPMs \cite{deStCroix_2020,GALLI2023167876}. The detector is housed in a cryostat located at McGill University, filled with $\sim$5 kg of LXe. Xenon gas from high pressure gas bottles is passed through a heated zirconium getter (MonoTorr PS3-MT3) and a cold SAES 902 inline-purifier before being condensed inside the LoLX cryostat. The cryostat is cooled by liquid nitrogen (LN$_{2}$), using an open-loop with liquid flow control, to reach the LXe condensation temperature of 165 K at $\sim$1 atm. Slow control and data acquisition is handled through custom C++ applications in the MIDAS framework \cite{ritt2019midas}. 

\begin{figure}[ht]
    \centering
    \includegraphics[width=0.7\linewidth]{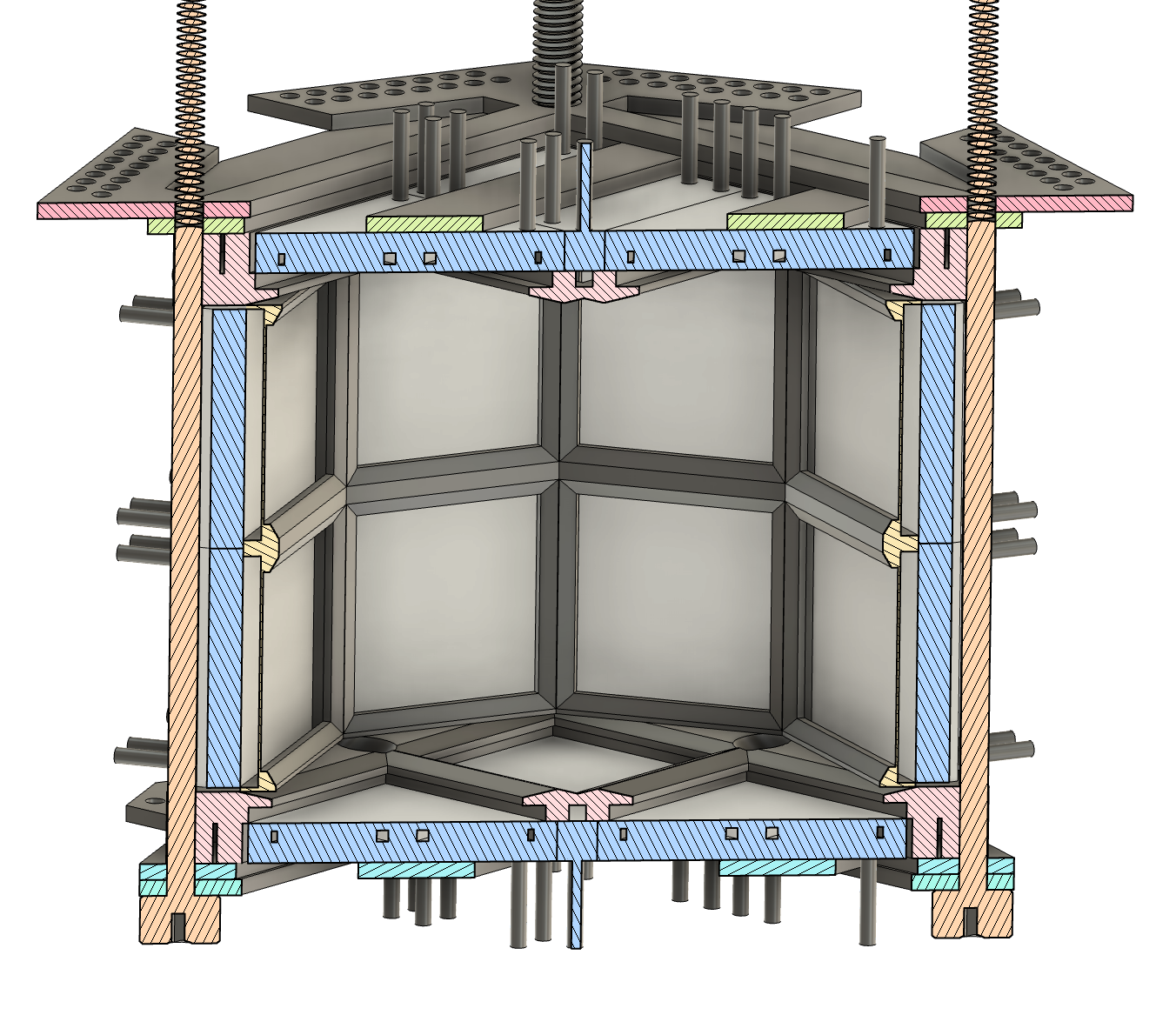}
    \caption{Cross-section CAD rendering of the LoLX detector. SiPM packages of 4 held in place by the resin 3D printed cage. 23 packages out of the 24 are placed behind optical filters.}
    \label{fig:lolx-det}
\end{figure}

\begin{figure}[ht]
    \centering
    \includegraphics[width=0.95\linewidth]{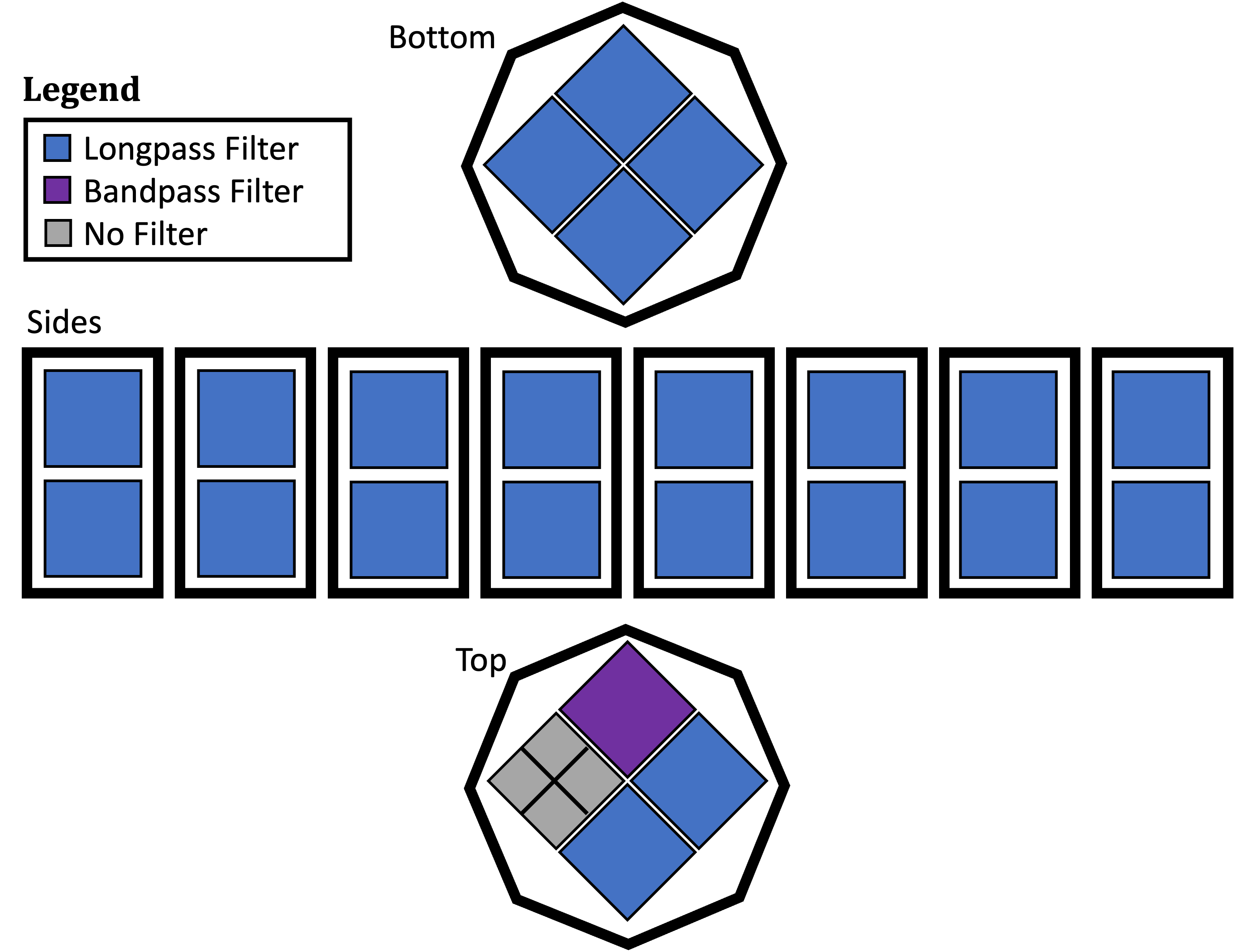}
    \caption{Illustration of filter configuration and layout of the LoLX detector. The orientation of the top/bottom are inverted for consistency.}
    \label{fig:lolx-unwrapped}
\end{figure}

The LoLX detector includes 96 Hamamatsu VUV4 SiPMs, in 24 packages of 4, configured in an octagonal prism geometry, shown in cross-section rendering in \autoref{fig:lolx-det}. Of the 24 SiPM packages, 23 are fitted with optical filters for wavelength selection of incident light, illustrated in \autoref{fig:lolx-unwrapped}. The SiPM packages are housed in a stereolithographic 3D printed acrylate polymer, using `Durable Resin' from FormLabs \cite{formlabs}. SiPM signals are transmitted through high-density coaxial feedthroughs to a custom amplifier board using high bandwidth RF amplifiers, which also supply reverse-bias voltages for the SiPMs. Each amplifier is controlled by a single-board computers (NanoPis) allowing for control of individual channels and serve as remote modules to the main data acquisition (DAQ) computer, communicating through the MIDAS server with the central database. The amplified signal is read out by a CAEN V1740 digitizer with a sampling rate of 62.5 MS/s and 12-bit vertical resolution with a 2~V dynamic range. Before data-taking, the breakdown voltage of each SiPM is measured by reverse-IV curves using a visible pulsed LED source to ensure sufficient current for readout. This is necessary as the dark count rate at 170 K is $\mathcal{O}$(Hz) and the amplifier board's limit for current readback is approximately 1 nA. The pulse-finding algorithm extracts in real-time the single photoelectron (SPE) charge and amplitude from the waveforms recorded for each channel. By fitting Gaussian templates to the distributions of both charge and amplitude, the average SPE charge and amplitude for a single avalanche are calculated,  and used for online data-quality monitoring.

LoLX's 24 SiPM packages are read out as 30 total channels. The readout channels are distributed as follows: 

\begin{itemize}
    \item 22 VUV4 packages covered by longpass filters (LP) with a cutoff wavelength of 225 nm. Each group of 4 is actively summed at the preamplifier and digitized on a single channel.
    \item 1 VUV4 package with a bandpass  filter (BP) selecting LXe scintillation light. The four signals are digitized on individual channels.
    \item 1 VUV4 package with no optical filter (referred to as "bare" or "unfiltered"). The four signals are digitized on individual channels.
\end{itemize}


\begin{figure}[ht]
    \centering
    \includegraphics[width=0.95\linewidth]{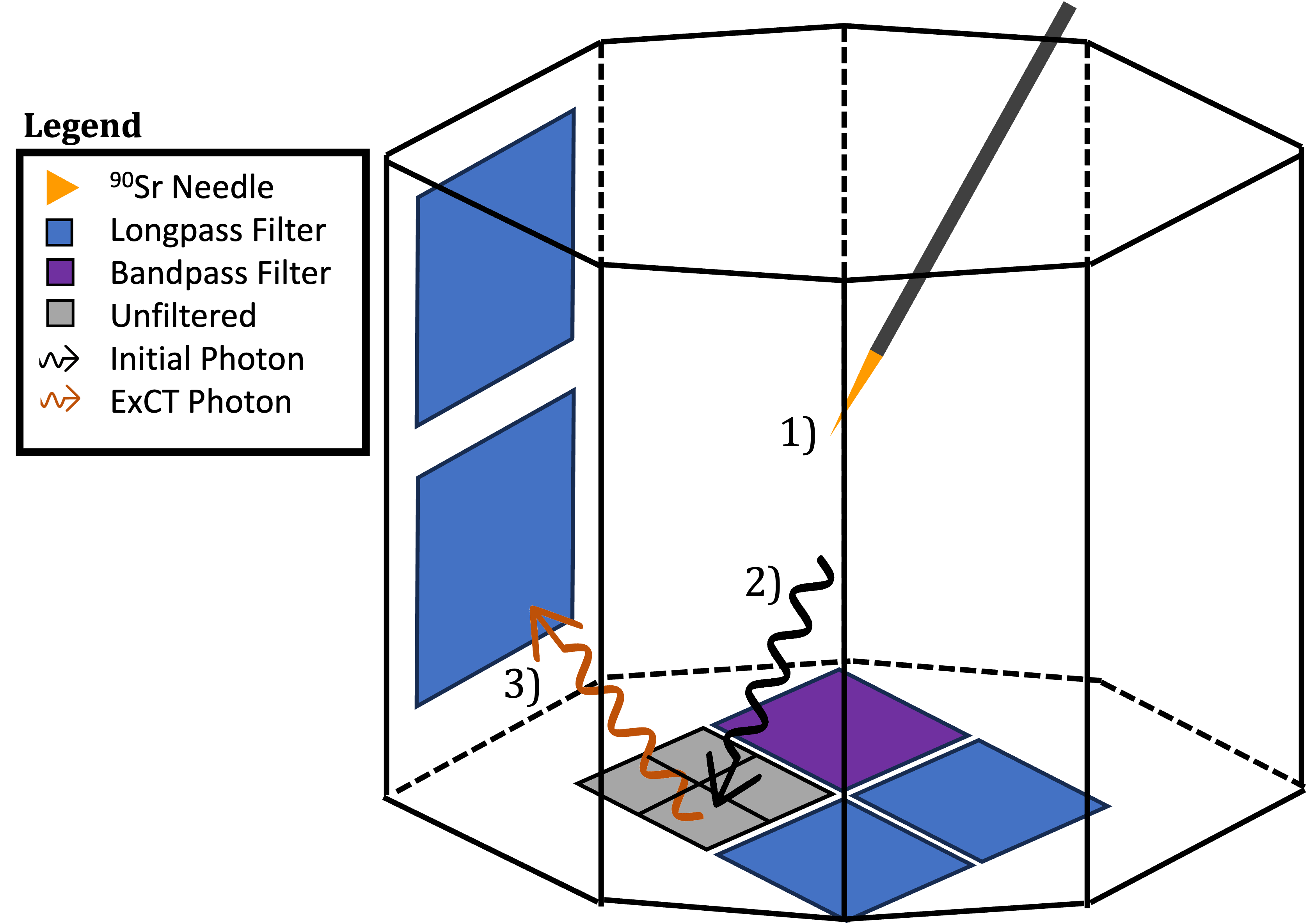}
    \caption{Illustration of initial photon detection and ExCT production in LoLX. After initial scintillation generated  from the energy deposition in LXe by the source, near-IR ExCT photons are produced in SiPM avalanches and can escape the device and transmit across to surrounding channels. This illustration is upside-down relative to the installed and operational orientation.}
    \label{fig:lolx-ext-ortho}
\end{figure}

The LoLX detector uses a $^{90}\text{Sr}$ beta source needle inserted from the bottom into the middle of the instrumented LXe volume. $^{90}\text{Sr}$ was selected to cover a wide range of energies as the daughter isotope $^{90}\text{Y}$ decays to the stable $^{90}\text{Zr}$ with an endpoint energy of 2278 keV \cite{BASU20201}.  \autoref{fig:lolx-ext-ortho} shows an illustration of the inverted detector together with an ExCT event. In stage 1) the primary scintillation occurs near the $^{90}$Sr source, in stage 2) a `seed' photon is detected by the bare SiPM, in stage 3) the avalanche produces a NIR photon that is detected by a nearby LP-filtered SiPM. The LP-filtered SiPMs constitute the majority of the sensitive surface area of the LoLX detector; in this work, they are the primary search channel for the detection of SiPM ExCT.
The optical properties of the LP filter and responsivity of SiPMs are detailed in the following section and illustrated in \autoref{fig:lolx-optical}.

\section{External cross-talk measurement procedure and simulation}
\label{sec:2}
This section presents an outline of the measurement procedure, event selection and analysis details for the SiPM external cross-talk study. This is followed by an outline of the simulation and mathematical framework to extract detector independent ExCT properties. The data were collected during a 4-day LXe run in October 2021, collected with two different applied overvoltages of 4 V and 5 V over multiple runs. Pulse-finding and waveform analysis is carried out on all channels for each event, saving all pulse and sub-pulse information for each run. SPE calibrations were performed and corresponding analysis thresholds set on a run-by-run basis, which corrects for the minor temperature induced gain variations from run to run.

\subsection{Data analysis}
\subsubsection{Event selection}
Data were taken with a >3 photon equivalent ADC threshold on any of the 4 channels in the unfiltered package. Events selected for analysis must pass a selection of data cleaning and analysis cuts:
\begin{itemize}
    \item No channels may saturate the digitizer's dynamic range (removes muon-like events with much higher light output than source events) 
    \item Stable pre-trigger baselines (removes events with saturation or digitizer clipping in the preceding event)
    \item Only one trigger candidate per event (removes accidental pile-up).
    \item Well-defined start time for the event (no significant pre-trigger light).
\end{itemize}

\begin{figure}[ht]
    \centering
    \includegraphics[width=0.9\linewidth]{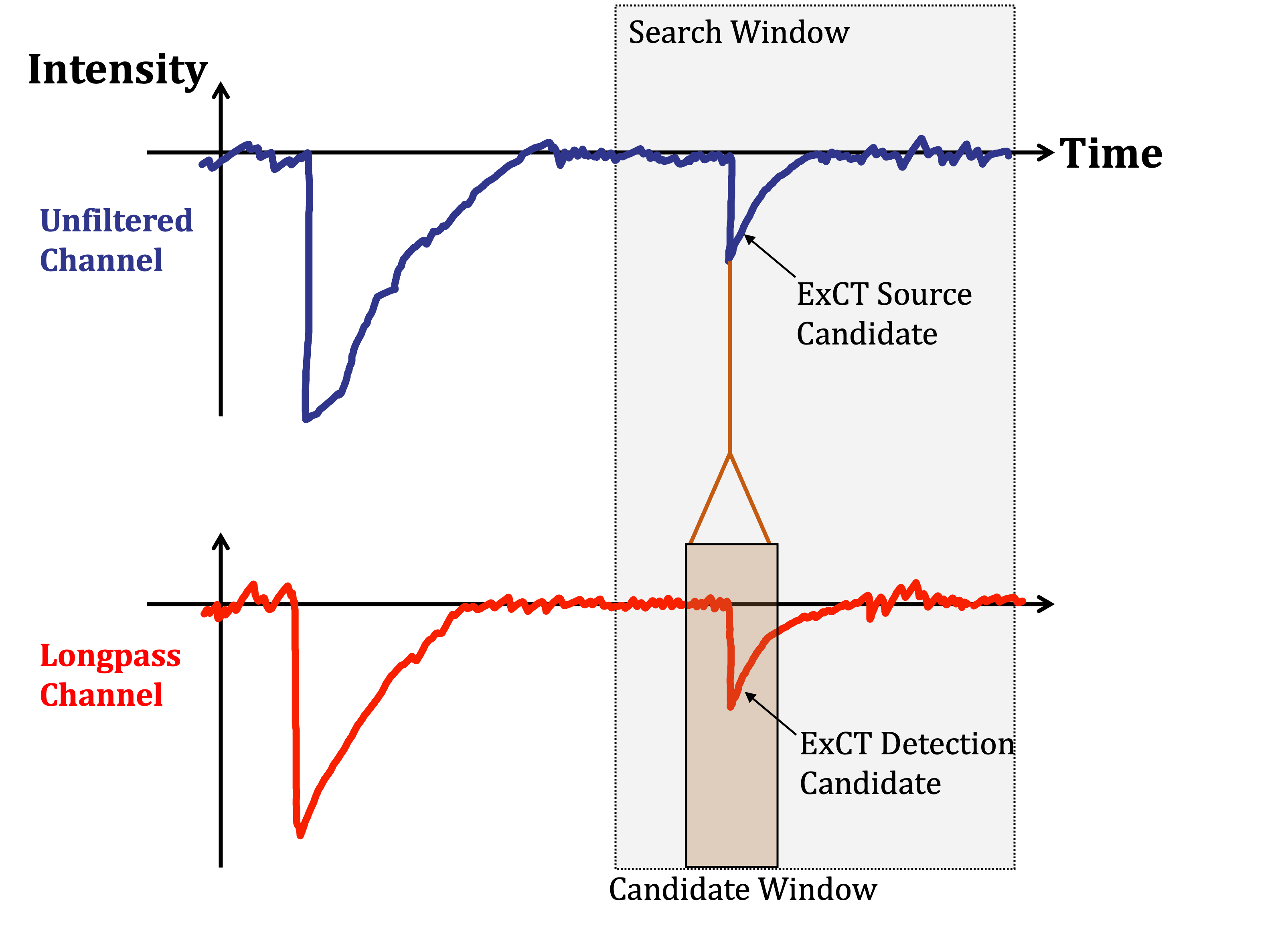}
    \caption{Illustration of a candidate ExCT source pulse and detection pulse. The time difference between pulses that are within the coincidence window is recorded for all pairings between the unfiltered and longpass channels. Within LoLX the transport time expected for SiPM ExCT is 3 orders of magnitude faster than the digitizer sampling rate: the coincidence signal is expected to appear with approximately zero time difference.}
    \label{fig:lolx-ext-method}
\end{figure}

\subsubsection{Analysis procedure}
\label{sec:data_analysis}

The analysis is restricted to a search for ExCT between bare and LP-filtered SiPMs, where the goal is to study the excess of time-coincident pulses between these channel groups. This is achieved by examining the distribution of the time difference between pulses across each channel group, as ExCT is expected to produce a prompt coincidence signal between two SiPMs, as shown in~\autoref{fig:lolx-ext-method}. The bare-to-filtered process was selected for two reasons. It maximizes the statistics of successful ExCT events available for analysis, as the LP channels have a high solid-angle coverage relative to the bare SiPMs. Additionally, the LP-filtered channels have a smaller fraction of uncorrelated backgrounds due to the optical filtering, which blocks the majority of VUV photons that would otherwise produce delayed afterpulsing in the search channel. 

\begin{figure}[ht]
    \centering
    \includegraphics[width=0.9\linewidth]{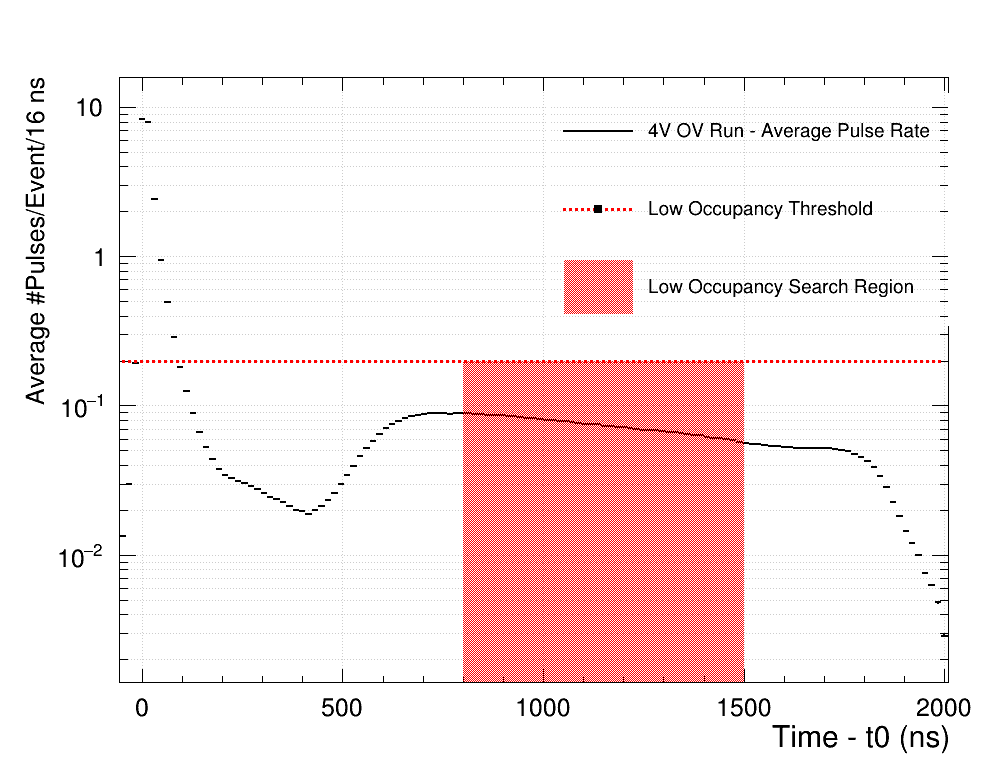}
    \caption{Occupancy of all SiPM channels inside LoLX, normalized to the number of events in 16 ns time bins for the 4~V OV dataset. The low-occupancy window extends from 800--1500 ns after the event start time. The dip in the distribution below 600 ns is an artifact of the pulse-finding convention selecting for primary pulses only, and not subpeaks. Due to the large initial scintillation signal, most subsequent avalanches are subpeaks to the primary scintillation light. }
    \label{fig:res-occupancy}
\end{figure}

As the first step in the analysis, each bare channel is searched for ExCT candidate pulses which must satisfy the following conditions:
\begin{itemize}
    \item Candidate pulses occur within the low occupancy search window, defined as the time window containing less than 0.2 pulses per event, on average. The search window is illustrated by the red band in \autoref{fig:res-occupancy}.
    \item Candidate pulses are SPE-like from charge-based estimation, to ensure approximately constant emission intensity among source pulses.
    \item Candidate pulses have no additional subpeaks (no afterpulsing).
\end{itemize}

To determine the ``low-occupancy" window region, the average number of pulses for all channels was measured as a function of time in the event window. The distribution of average number of pulses per 16 ns bin is labelled `occupancy' and is shown in \autoref{fig:res-occupancy}, demonstrating that the low occupancy requirements listed above are met across the entire search window. This requirement reduces uncorrelated coincidences between channels.

\begin{figure}[ht]
    \centering
    \includegraphics[width=0.95\linewidth]{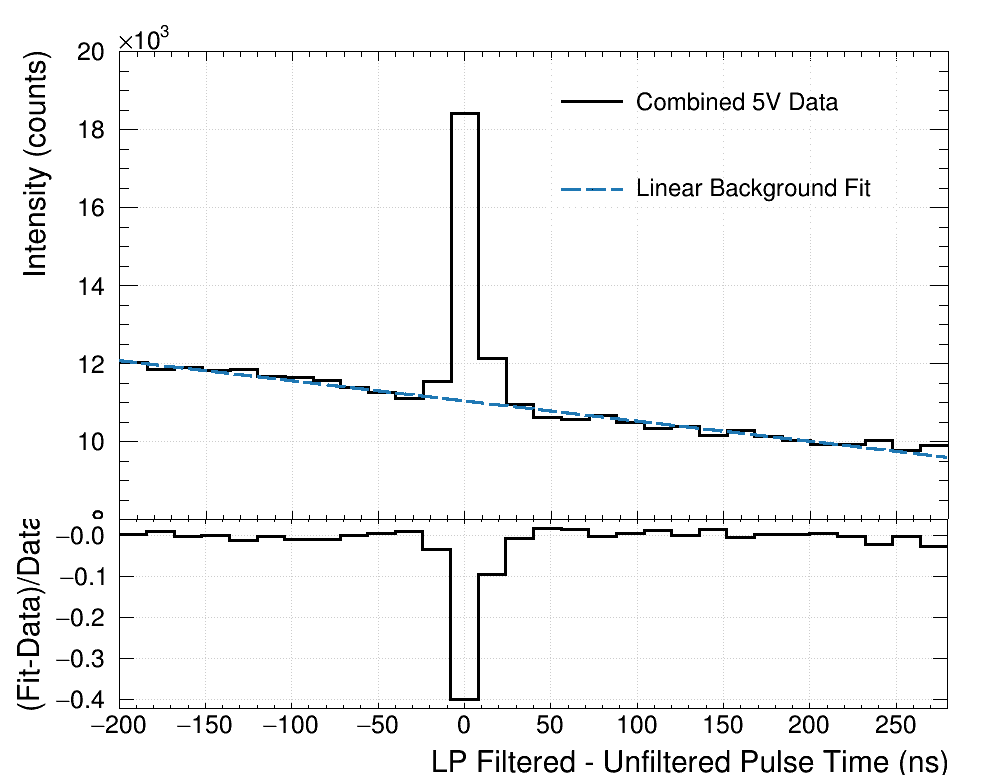}
    \caption{Fit to the delta-time distribution for all 5~V data summed. The linear contribution to the combined fit is shown and used in subsequent analysis.}
    \label{fig:sample-fit}
\end{figure}

A relative search window of $\pm296$ ns is defined with respect to the bare pulse's leading edge, as shown in \autoref{fig:lolx-ext-method}. This relative search window is much larger than the $\mathcal{O}(10$~ns) timing resolution of LoLX and includes both random and coincident events. All LP-filtered channels are scanned within the relative window for SPE-like candidate pulses. The time difference $\Delta t$ between the leading edge of the bare and any candidate LP pulse is calculated and histogrammed. The pulse pairings are summed over the four bare channels and all LP-filtered channels. These $\Delta t$ distributions are produced per run and combined for each of the two overvoltage values: an example is shown in \autoref{fig:sample-fit}. The bin size of 16~ns is set by the digitization speed (62.5~MHz, or 16~ns samples), pulse-finder resolution and DAQ jitter.

\par 
This $\Delta t$ distribution has two main features: a peak of correlated detections centred at $\Delta t$ = 0, and a falling uncorrelated background from random coincidence events.  The background component outside the peak region is fit to a first order polynomial, the results of which are used to constrain the starting parameters of a combined linear background + Gaussian signal fit. The mean of the Gaussian fit is fixed to zero. An example of the fit is included in \autoref{fig:sample-fit}. Using the $\Delta t$ distribution, the `bare SiPM correlated pulse probability' between the bare and LP-filtered channels is given by:
\begin{equation}
    \label{eqn:ext-prob}
     P_{corr,B} = \frac{N_{correlated} - N_{background}}{N_B} ~.
\end{equation}

In this equation, $N_{B}$ is the number of total bare SiPM candidate pulses. The numerator gives the excess number of hits about $\Delta t = 0$, where $N_{correlated}$ is the integrated count under the Gaussian fit. $N_{background}$ is estimated by evaluating the linear component of the combined fit at the bin centres of the three signal bins, which span [-24 ns, 24 ns]. This timeframe is referred to as the `correlation' or `signal' window. This excess of correlated hits over the combinatoric background, centred at $\Delta t = 0$, is ascribed to external cross-talk. The temporal width of this peak is attributed to the timing resolution of the LoLX signal chain and pulse analysis.

For a given SPE-like pulse in a bare SiPM, $P_{corr,B}$ is the probability of observing a correlated and time-coincident SPE-like pulse in any LP channel. Due to the normalization in \autoref{eqn:ext-prob}, the values of $\pcb$ are not the probability of cross-talk from one SiPM to another: this is because $\pcb$ includes contributions from two processes. In the `forward' process (bare $\rightarrow$ LP), the bare SiPM is the true source of the ExCT photons, triggering one of the `target' LP SiPMs. Conversely, in the `backwards' process (bare $\leftarrow$ LP), the LP filtered SiPM is the true source of the ExCT photon triggering a bare SiPM. Candidate pulses in the bare SiPM which fail to trigger a coincident ExCT pulse are included in the normalization $N_B$, however failed backwards trials (SPE source pulses in the LP channel) are not. Due to the negligible transit time for ExCT photons across LoLX  it is impossible to resolve these two processes using timing information. Correcting for this `backward correlation' and its contribution to $\pcb$ is done in Section~\ref{sec:sim_analysis}, to yield a true `source-to-target' probability and the detector-independent photon emission intensity from silicon.

\subsection{External cross-talk simulation}
\label{sec:ext-sim}

\subsubsection{Simulation procedure and structure}
\label{sec:ext-sim-procedure}

\begin{table*}[!ht]
    \centering
    \caption{Table summarizing relevant parameters for external cross-talk simulation.}
    \begin{tabular}{l|l|l}
        \textbf{Simulation Property} & \textbf{Value(s)} & \textbf{Reference} \\
        \hline
        Hamamatsu SiPM Efficiency & See \autoref{fig:lolx-optical} & \cite{hpk_vuv4_ds} \\
        ExCT Emission Distribution (wavelength) & See \autoref{fig:lolx-optical} & Generated from \cite{raymond2024stimulated} \\
        ExCT Emission Transmission (wavelength and angle) &  See \autoref{fig:ext-emitted}& Generated from \cite{raymond2024stimulated} \footnotemark[1] \\
        Long-pass Filter Transmission & See \autoref{fig:lolx-optical} & \cite{newport_longpass_filters} \\
        Band-pass Filter Transmission & 20\% @ 178 nm & \cite{esource_optics} \\
        Specular Reflectivity of SiPMs & See \autoref{fig:lolx-optical} & Derived \\
        Diffuse Reflectivity of SiPMs & 18\% & \cite{diffuse_vuv} \\
        Diffuse Reflectivity of Cage Material & $\sim$95\% & Derived from Fresnel Relations \\
    \end{tabular}
    \label{tab:simdata}
\end{table*}

\begin{figure}[ht]
    \centering
    \includegraphics[width=0.95\linewidth]{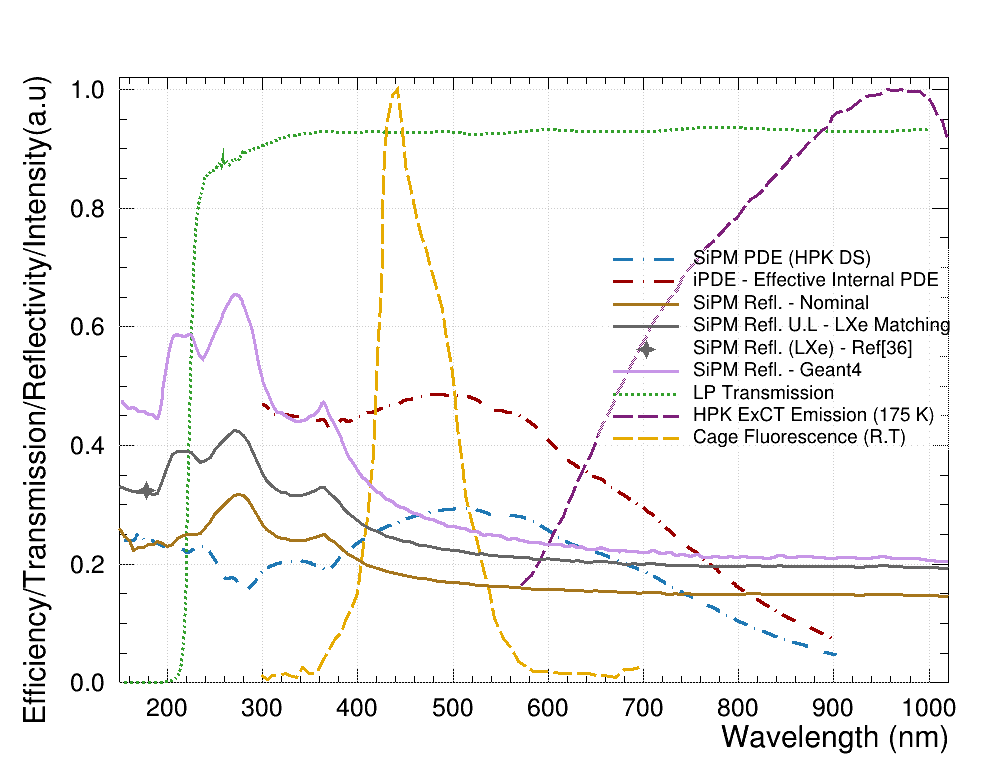}
    \caption{Optical response and input distributions for LoLX simulations. SiPM photo-detection efficiency data (PDE) are from the Hamamatsu data sheet, measured at room temperature and normal incidence. Fluorescence of the LoLX cage material was measured at ambient temperature at McGill University. See \autoref{tab:simdata} for references where applicable. The nominal and effective Geant4 reflectivity, systematic variations on the SiPM reflectivity (LXe U.L, described in Section \ref{sec:4}), and the internal PDE (iPDE) are shown for reference.}
    \label{fig:lolx-optical}
\end{figure}

\begin{figure}[ht]
    \centering
    \includegraphics[width=0.95\linewidth]{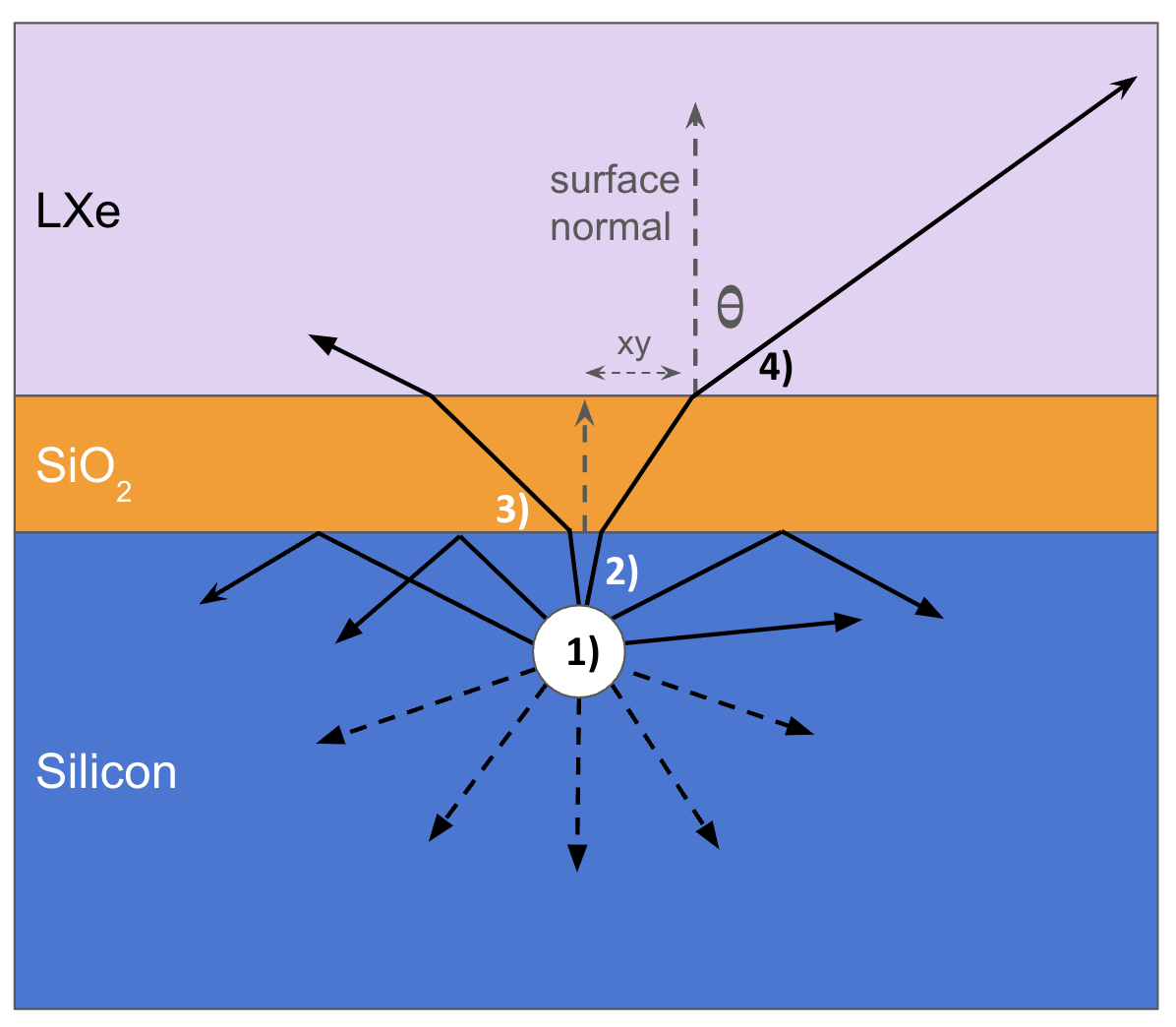}
    \caption{Illustration of SiPM ExCT and the associated simulation steps. Geometry and photon production location are not to scale. 1) The process begins at the detected photon position. 2) N ExCT photons are sampled from the wavelength distribution in \autoref{fig:lolx-optical} and produced isotropically. Only photons in the forward hemisphere, as defined by the SiPM surface normal, are actually simulated. 3) Photon transmission probabilities for emission into LXe are sampled. The majority of photons are internally reflected. 4) Photons surviving the emission check are translated vertically into the LXe volume, with their angle $\theta$ calculated from Snell's law. This ignores the XY displacement due to refraction, which is on the $\mathcal{O}$(nm) scale and is negligible. The distribution of theta vs wavelength is shown in \autoref{fig:ext-emitted}, with a mode near 45$\degree$ due to the strong refraction out of silicon.}
    \label{fig:ext-zoom}
\end{figure}

LoLX simulations are carried out using a custom simulation package written using the Geant4 framework \cite{AGOSTINELLI2003250}, with NEST \cite{nest} integrated to generate the liquid xenon scintillation light (using 10~V/cm as the zero-field approximation). The LoLX model uses a surface-boundary based detection paradigm for the optical photons. All relevant optical surfaces within the LoLX detector are defined in the simulation with properties applied over the wavelength range from 150 nm to 1000 nm (see \autoref{fig:lolx-optical} and \autoref{tab:simdata} for reference). Custom physics and optical boundary classes are implemented to produce ExCT photons directly in the simulation, similar to a fluorescence process. The ExCT process, simulation inputs and assumptions are as follows:

\begin{enumerate}
    \item The wavelength dependent ExCT emission spectrum, characterized ex-situ \cite{raymond2024stimulated}, is loaded into the custom physics process, alongside look-up tables for all relevant optical properties inside LoLX.
    \item Photons that pass through an optical boundary to a detector surface and are successfully detected trigger the ExCT process, using a modified version of G4OpBoundaryPhysics.
    \item Photons are produced within the bulk silicon, with a tuneable mean number of photons per avalanche. Those photons are produced isotropically in the outgoing hemisphere, with a wavelength sampled from the emission curve in \autoref{fig:lolx-optical}. To exit the device, photons must pass a transmission or `emission' check from the input optical model outlined below. This process is illustrated in \autoref{fig:ext-zoom}.
    \item ExCT photons passing the emission check are manually refracted and placed into the LXe directly in-front of the SiPM volume. This is to avoid built-in optical photon transport through the SiPM volumes, which was already accomplished in the previous step. These photons, now within the LXe volume, are flagged as ExCT photons for future analysis. ExCT photons produced in the simulation are added to the stack of secondary particles and transported through the LoLX Geant4 simulation. They may induce further ExCT when subsequently detected.
\end{enumerate}

The sampled ExCT wavelength spectrum is the spectrum \textit{inside} the silicon; it is the vacuum spectrum measured in~\cite{raymond2024stimulated} corrected for the transmission from silicon to vacuum\footnotemark[1]. The estimated transmission for ExCT photons from silicon into LXe is also provided by \cite{raymond2024stimulated}. This two-dimensional transmission information is used as the input optical model  for the ExCT photon emission into LXe. Due to the large refractive index mismatch between silicon, silicon dioxide, and LXe, the majority of the ExCT photons are internally reflected. Internally reflected ExCT photons are killed in the simulation to avoid double-counting direct optical cross-talk.

\footnotetext[1]{Produced for this study by the authors of \cite{raymond2024stimulated}. Based on ANSYS Lumerical simulations of light emission from HPK SiPM into LXe and/or vacuum, for a thin quartz film between the silicon and active medium.}

\begin{figure}[ht]
    \centering
    \includegraphics[width=0.9\linewidth]{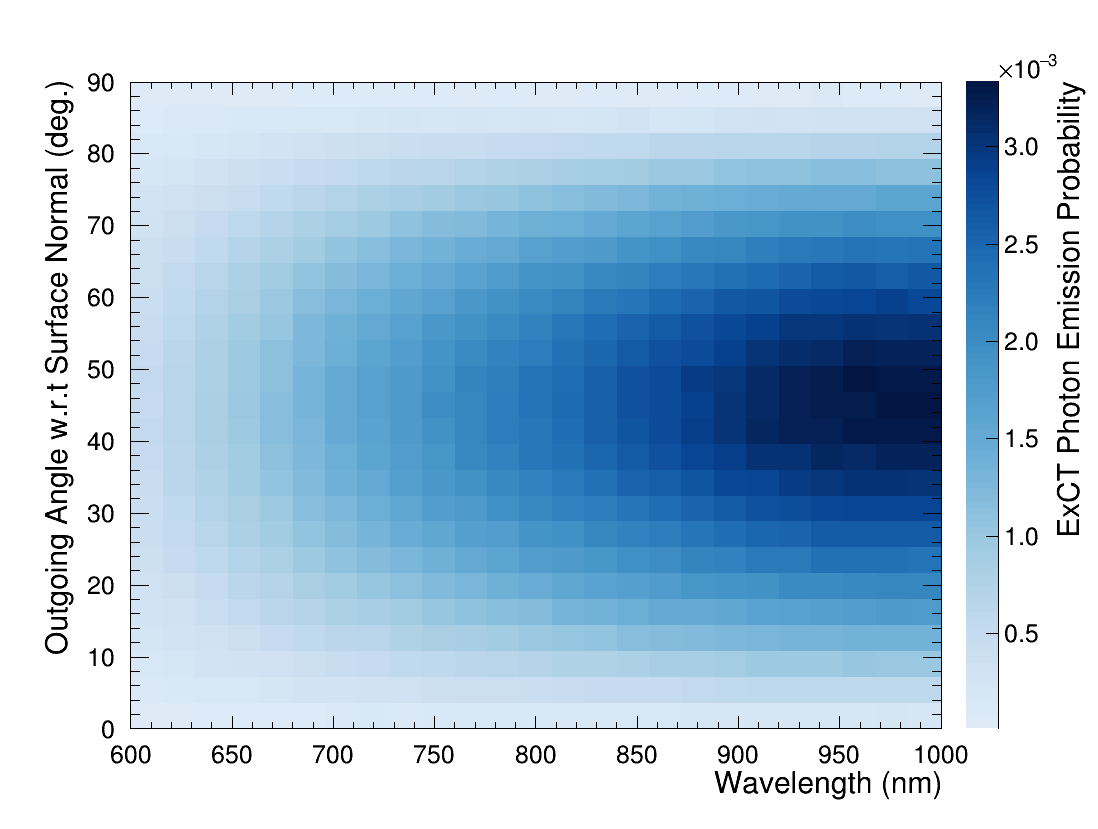}
    \caption{Angular distribution of emitted photons into LXe, calculated from the procedure detailed in Section~\ref{sec:ext-sim-procedure}. Probabilities are normalized to the proportion of emitted photons in the outgoing hemisphere ($2\pi$).}
    \label{fig:ext-emitted}
\end{figure}

The resulting distribution of outgoing ExCT photons from this model is shown in \autoref{fig:ext-emitted}, with a broad emission centred at $45\degree$ with respect to the SiPM surface normal. While photon emission within the silicon is assumed to be isotropic, this oblique resulting emission into LXe is due to the combination of strong refraction at the silicon-LXe interface, and total internal reflection reducing the intensity at very oblique angles. At the visible wavelengths of ExCT, total internal reflection for photons in the silicon occurs at approximately $20\degree$. 

Having implemented the ExCT process directly into the simulation framework, its effects are automatically included in subsequent LoLX simulation runs, enabling an estimate of the average transport and detection efficiency for ExCT photons across LoLX.

\subsubsection{Simulated SiPM optics for incoming photons}
\label{sec:sipm-optics}

\begin{figure}[ht]
    \centering
    \includegraphics[width=0.99\linewidth]{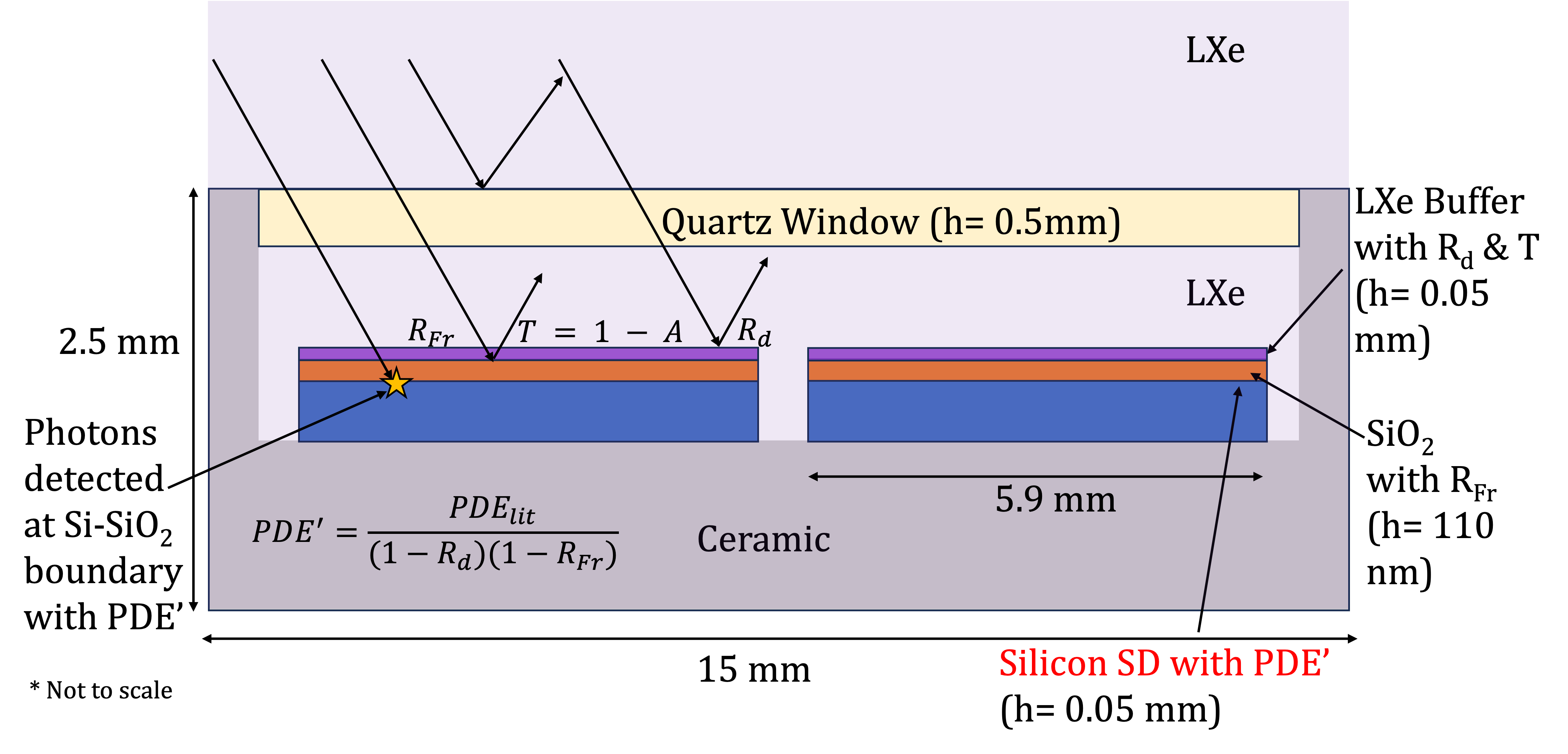}
    \caption{Illustration of HPK VUV4 package (cross-section) and optical boundaries in the simulation. To circumvent restrictions in optics from Geant4, a secondary optical boundary (LXe Buffer) is implemented with directional surface optical properties to match literature values for detected and reflected photons.}
    \label{fig:sipm-optics}
\end{figure}

The optical simulation of photons incident on the SiPM surface from LXe is described in this section. As both the SiPM PDE and reflectivity are strongly coupled to the ExCT measurement, a detailed simulation strategy is required. The method described below provides proper Fresnel coupling to the LXe medium (wavelength and angular dependence) using available $n, k$ optical data. Directional optical surfaces are used to scale the total reflectivity to match literature values for vacuum and LXe~\cite{nexo_reflectivity_vacuum, Wagenpfeil_2021}. An internal SiPM PDE is defined, factoring out the optics to achieve a net PDE that agrees with the HPK datasheet \cite{hpk_vuv4_ds} for different reflectivity values. This simulation method, including relevant optical surfaces in the simulation model, is illustrated in \autoref{fig:sipm-optics}. 

Prior to reaching the physical SiPM surface, incoming photons of any wavelength are diffusely reflected off the first directional surface with probability $R_{d}\approx 18\%$, in agreement with \cite{nexo_reflectivity_vacuum}. The diffuse reflectivity is mainly attributed to the SiPM's surface microstructure. For this application, the SiPMs are simulated as a flat silicon-SiO$_{2}$-medium stack, with Fresnel ray optics at each interface. The $n$ and $k$ input data for silicon are taken from \cite{nSi_aspnes} and \cite{nSi_pierce}. Ref.~\cite{nSi_pierce} is for amorphous silicon which was only measured in the VUV and IR regions, where it matches and extends from the crystalline Si data of \cite{nSi_aspnes}. The SiO$_{2}$ data is taken from~\cite{lithography} and~\cite{Malitson}. The total Fresnel specular reflectivity of these layers is labeled as $R_{Fr}$. By using ray optics, possible thin-film interference effects are neglected, although the shape of the $R_{Fr}$ curve is similar to the vacuum reflectivity, lowered by a constant vertical scaling. The higher $R_{Fr}$ is partially attributable to the absence of insensitive, non-specular SiPM surfaces. To reduce the specular reflection in the simulation, excess reflected photons are absorbed by a second directional surface with probability $A$. The absorption parameter $A$ is an effective parameter, tuned to control the total reflectivity. Physically, $A$ is related to absorption on the SiPM microstructure, absorption on SiO$_2$, and any other effects which may reduce specular reflection from the Fresnel predicted value.

After the diffuse and specular reflectivity, a photon has a probability of $(1-R_d)(1-R_{Fr})$ of being transmitted into the bulk. Therefore, the Hamamatsu provided PDE \cite{hpk_vuv4_ds}, labelled $PDE_{HPK}$, is normalized by this transmission yielding an internal efficiency $iPDE$. The normalization is calculated using $R_{Fr}$ in vacuum at normal incidence, matching the conditions under which $PDE_{HPK}$ was measured. The internal PDE treats all photons equivalently once inside the silicon, regardless of their angle. Due to the strong refraction into silicon, this approximation is reasonable. A detected photon will pass the trials $(1-R_d), (1-R_{Fr}), iPDE$, with $iPDE$ defined as below:
\begin{equation}
    iPDE = \frac{PDE_{HPK}}{(1-R_d)(1-R_{Fr})} 
\end{equation}

A specularly reflected photon must sequentially pass the Bernoulli trials $(1-R_d), R_{Fr}, (1-A)$. For the nominal estimate of the SiPM reflectivity, the $A$ parameter is tuned such that $R_{spec}$ is equal to the vacuum reflectivity $R_{lit}$ as measured in \cite{nexo_reflectivity_vacuum} (using $R_{Fr}$ calculated in vacuum): 
\begin{equation}
    A_{nominal} = 1 - \frac{R_{lit}}{R_{Fr}(1-R_d)}
\end{equation}
The measurement in \cite{nexo_reflectivity_vacuum} extends to a wavelength of 280~nm. For larger wavelengths, $A$ is fixed at the 280~nm value of $0.45$. Beyond 600~nm (relevant to the ExCT measurement) this results in a fairly flat specular reflectivity of $\sim10\%$ in vacuum or $\sim 15\%$ in LXe. This nominal estimate is also taken as the lower bound on SiPM reflectivity, as the $R_{lit}$ values from \cite{nexo_reflectivity_vacuum} are lower than the expectation of the sensitive area (fill factor or $FF$) multiplied by the Fresnel reflectivity, $FF \cdot R_{Fr}$, and absorption is expected to decrease from 280~nm to the visible.

To estimate a systematic upper-limit for the reflectivity in the NIR, the diffuse reflectivity is assumed to be due to the SiPM microstructure (a relative contribution of $1-FF$), while all the active area (a contribution of $FF$) is assumed to be purely specular with $R_{Fr}$. To achieve this, the effective specular reflectivity $(1-R_d)R_{Fr}(1-A)$ is equated to the intended physical reflectivity, $FF \cdot R_{Fr}$, and rearranged for $A$:
\begin{equation}
    A = 1-\frac{FF}{1-R_D} = 1-\frac{0.6}{1-0.18} = 0.27~.
\end{equation}
At 600 nm, this results in a specular reflectivity of roughly 16\% in vacuum and 20\% in LXe. At the LXe scintillation wavelength this approach gives a specular reflectivity of 33\%, in agreement with the values measured by the Munster group \cite{Wagenpfeil_2021} ranging in $28-34\%$, giving confidence in this strategy (\autoref{fig:lolx-optical}).

In summary, the conservative estimate for SiPM reflectivity agrees with \cite{nexo_reflectivity_vacuum} in vacuum and the upper limit agrees with \cite{Wagenpfeil_2021} in LXe at VUV wavelengths. In each case the effective PDE matches the curve provided by HPK, at normal incidence. This approach does not simulate any shadowing or angular dependence due to the SiPM's surface structure, however the geometric shadowing from the LoLX cage and SiPM packaging, included in the Geant4 simulation, limits incoming photons to roughly $\sim 65^\circ$ and should dominate over the effect of surface structure.

\subsubsection{Simulation analysis}
\label{sec:sim_analysis}
To differentiate between forward (bare $\rightarrow$ LP) and backwards (bare $\leftarrow$ LP) processes, two different ExCT simulations are carried out. For each simulation, mono-energetic beta decays are simulated at the centre of the detector to seed the ExCT process. In the first simulation, photon emission is exclusively simulated from the bare SiPMs, producing a map of ExCT transport and detection of the forward process. In the second simulation, ExCT photons are only emitted from the LP-filtered channels. Thus, the first simulation estimates the total average transport and detection probability of ExCT for the forwards process, while the second the probability for the backwards process.

Using the Monte-Carlo truth information for the physical source of each detected photon in the simulation, the ExCT transport and detection efficiency is simply the ratio of the detected ExCT photons on one set of channels to the number of emitted ExCT photons on the source channels which have been enabled in the simulation. These probabilities are hereafter referred to as $\pf$ and $\pb$ for the forward and backward scenarios, respectively.

\begin{equation}
\label{eq:mcprob}
    P_{\rightarrow} = \frac{1}{N_{\gamma}}\sum_{\substack{i \in LP}}{N_{ExCT, i}} \\,\\
    P_{\leftarrow} =  \frac{1}{N_{\gamma}}\sum_{\substack{i \in Bare}}{N_{ExCT, i}} 
\end{equation}
Here $N_{\gamma}$ is the recorded total number of emitted ExCT photons in either simulation. The summation represents summing the number of detected ExCT photons in channel $i$ over all the channels of LP or bare SiPMs, respectively. The estimates of $\pf$ and $\pb$ are the averaged transport and detection probability for ExCT photons between the bare and LP filtered channel groups. This probability is a detector-specific transport property, and is not affected by the energy or type of simulated event. For the backwards process, all LP channels may generate ExCT photons, so $\pb$ includes higher-order contributions where one LP ExCT event may trigger a second LP channel, which triggers the bare SiPM.

DiCT is not included in the simulations of $\pf$ or $\pb$. DiCT can enhance ExCT by producing additional avalanches. The mathematical model in Section~\ref{sec:eff-model} explicitly excludes DiCT so that its omission from the simulation of $\pf$ is valid. Within the SPE-enforced constraints of the effective model outlined below, DiCT can only contribute to $\pb$ in higher order cross-talk terms, which would populate 3 channels simultaneously. In the data analysis, the number of 3-channel-correlated events was negligible, within statistical uncertainty. It is concluded that the difference between the data sensitive $\pb$ and simulated value, arising from the omission of DiCT in simulation, can be neglected within the scope of the presented measurement.

A minimal version of the modified Boundary Process class, the ExCT Process class, detailed SiPM optics, and the required helper classes used in this work are available for community use as an extension of the Geant4 advanced example ``underground physics" \cite{dgallacher1:g4-ext}.

\subsection{Model for ExCT coincidence probability}
\label{sec:eff-model}

In this section, a framework is introduced to separate the forward (bare $\rightarrow$ LP) and backward ExCT contributions to the observable $\pcb$. Correcting for this `backwards correlation' allows the calculation of the detector-independent photon yield $\nxe$ from the data: $\nxe$ is the average number of ExCT photons emitted into LXe per avalanche and can be applied to other LXe experiments.

In the definition of $P_{corr,B}$ (\autoref{eqn:ext-prob}) the numerator gives the sum of all successful ExCT events: coincident pulse pairs between the LP filtered and bare SiPMs. However, the denominator is the sum of all SPE-like pulses in the bare SiPM \textit{only}. Thus, the normalization accounts for trial pulses for the forward process (bare $\rightarrow$ LP), while the numerator is the sum of successful ExCT events initiated by trial pulses in either channel (both the forwards and backwards processes).

The strategy for this correction is to write the observable $P_{corr,B}$ as a function of the previously defined transport probabilities $\pf$ and $\pb$, and the photon yield $\nxe$. The following framework describes ExCT in LoLX to first order only; negligible corrections for cross-talk of cross-talk are not included. This framework may be applied to other experiments, provided the detector-dependent $\pf$ and $\pb$ can be estimated. 

The probability of ExCT detection scales with the photon yield $\nxe$. For a single ExCT success, this probability is the sum of Poisson processes generating $N_\gamma$ photons with an average of $\nxe$, with each Poisson term multiplied by the binomial probability of a single success for $N_\gamma$ trials and probability of success $\pf$ or  $\pb$. For a forward ExCT event the total probability $P_{Forw}$ can be written as: 
\begin{align}
    \label{eq:pforward_approx}
    P_{Forw} = \sum_{k=1}^\infty (& Poisson(k=N_\gamma, \lambda=N_{Xe})\cdot \nonumber \\
    & Binomial(n=N_\gamma, k=1, p=\pf))~;
\end{align}
\begin{align}
    P_{Forw} &= \sum_{N_\gamma=1}^\infty\frac{N_{Xe}^{N_\gamma}e^{-N_{Xe}}}{{N_\gamma}!} N_\gamma\pf (1-\pf)^{N_\gamma-1} \approx N_{Xe} \pf ~.   
\end{align}
The approximation $P_{Forw} \approx N_{Xe} \pf $ is accurate to within 3\% relative, for the small $N_{Xe}$ and $\pf, \pb$ values relevant for this study (see \autoref{tab:modelinputs}). Similarly, for the backwards process $P_{Back} \approx \nxe\pb$.

The probability of DiCT within a SiPM is denoted $\pix$. $N_{1,B}$ and $N_{1,LP}$ are the number of \textit{source} or unperturbed single PE pulses in the bare or LP channels, unperturbed by the ExCT process. Physically these pulses are the sum of dark noise, very late after pulses, stray light and photoluminescence. The numerator in \autoref{eqn:ext-prob} will be denoted as $N_{corr,B}$, representing the number of correlated, SPE-like, ExCT induced events across channel groups. This includes contributions from both processes.
\begin{align}
    N_{corr,B} & = (P_{Forw}N_{1,B} + P_{Back}N_{1,LP})(1-P_{DiCT}) \\
    N_{corr,B} & \approx (N_{Xe}P_{\rightarrow}N_{1,B} + N_{Xe}P_{\leftarrow}N_{1,LP})(1-P_{DiCT})
    \label{eq:N_cb_mathModel}
\end{align}

\noindent where $P_{Forw}$ and $P_{Back}$ are the exact Poisson-binomial sums given in \autoref{eq:pforward_approx}. This equation independently relates the contributions from the forward and backward processes to the observable. The factor of $1-\pix$ ensures the analysis condition that resulting pulses are SPE-like. This model does not account for `runaway' ExCT, meaning cross-talk inducing additional cross-talk. However, for the moderate voltages used in LoLX this runaway contribution is negligible and is thus ignored\footnote{To explicitly exclude runaway, the probability of no additional events must be included. For the forward term, the correction factor is approximately $1-\nxe \pb (1-\pix)$, which ensures the resulting avalanche in the LP SiPM does not trigger a subsequent bare avalanche, as the bare SiPM would then fail the SPE analysis condition. The correction for the backwards term is even smaller, $1 - \nxe\pf(1-\pix)\nxe\pb$ which is a third order correction. These higher order corrections are less than 1\% and are neglected.}.

To remain consistent with the normalization of $P_{corr,B}$ (\autoref{eqn:ext-prob}), \autoref{eq:N_cb_mathModel} is divided by the total number of SPE-like pulses in the bare SiPM group. This includes the unperturbed number of pulses $N_{1,B}$ and the ExCT events initiated by the LP channels, the backwards process. $P_{corr,B}$ then becomes:
\begin{equation*}
    P_{corr,B} = \frac{(P_{Forw}N_{1,B} + P_{Back}N_{1,LP})(1-P_{DiCT})}{N_{1,B} + P_{Back}N_{1,LP}(1-P_{DiCT})} ~.
\end{equation*}
Defining $R_{R}$ as the ratio of the unperturbed rates in either channel, $R_{R} = R_{1,LP}/R_{1,B} = N_{1,LP}/N_{1,B}$, the equation can be simplified to arrive at the following expressions.
\begin{equation}
\label{eq:effective-model}
    P_{corr,B} = \frac{(P_{Forw} + P_{Back}R_R)}{\frac{1}{(1-\pix)} + P_{Back}R_R} \approx \frac{\pf + R_R\pb}{\frac{1}{\nxe (1-\pix)} +R_R \pb}
\end{equation}
The experimentally observed rate ratio is denoted $\drr$, which differs from the unperturbed or ExCT-free ratio $R_{R}$. Similar to deriving $\pcb$ the experimentally observed ratio can be written using the unperturbed rates plus the ExCT contribution and the function $R_R(\drr)$ is obtained as follows:
\begin{align}
    \drr &= \frac{R_{1, LP} + R_{1,B}P_{Forw}(1-\pix)}{R_{1, B} + R_{1,LP}P_{Back}(1-\pix)} ~;
    \label{eq:rate-ratio-raw}
    \\
    R_R &=\frac{\drr - P_{Forw}(1-\pix)}{1 - \drr P_{Back}(1-\pix)} \approx \frac{\drr - N_{Xe}\pf(1-\pix)}{1 - \drr N_{Xe}\pb(1-\pix)}~.
    \label{eq:rate-ratio}
\end{align}
Again, the runaway correction terms neglected in \autoref{eq:N_cb_mathModel} are ignored here. Naturally, for very weak ExCT, $\nxe \rightarrow 0$, the experimental rate ratio $\drr$ approaches the unperturbed ratio $R_{R}$.

\autoref{eq:effective-model} and \autoref{eq:rate-ratio} form a set of nested equations for $\pcb$ as a function of photon yield $\nxe$ and other inputs. The inputs $\pf, \pb$ are evaluated from simulation, $\drr$ from data, and $\pix$ from literature. These equations are solved numerically to determine the value of $\nxe$ that gives agreement with the experimentally measured $\pcb$.

\section{Results}
\label{sec:3}
\subsection{Delta-time distribution and correlated pulse probability}
The time difference $\Delta t$ between SPE pulses on the bare and LP channels is calculated to estimate the rate of correlated and uncorrelated events. The distribution is shown in \autoref{fig:sample-fit}. This distribution's peak about $\Delta t = 0$ is ascribed to ExCT, with the peak width driven by the timing resolution of LoLX. Based on the time difference $\Delta t$ between SPE pulses on the bare and LP channels showin in~\autoref{fig:sample-fit}, the ExCT signal excess above background is evaluated following the fitting procedure described in Section \ref{sec:data_analysis}. The number of candidate bare SiPM pulses $N_B$ is used to normalize this excess, yielding the correlated pulse probabilities $\pcb = 3.8\%$ and $5.2\%$ at 4~V and 5~V overvoltage, respectively. As previously stated, these correlated pulse probabilities include contributions from both the forward and backward processes. \autoref{eq:rate-ratio-raw} can be examined to show how $\pcb$ is not directly proportional to the ExCT probability. For very large $R_R$, meaning a high SPE pulse rate in the LP channels relative to the bare channels, $\pcb$ approaches 1 even for small $\nxe$, i.e. weak ExCT. Thus, the model outlined in Section \ref{sec:eff-model} must be used in conjunction with the simulated transport efficiencies $\pf$ and $\pb$ and the measured $\pcb$ to extract the ExCT probability. After the discussion of systematic uncertainties in Section~\ref{sec:syst_unc}, the resulting ExCT probabilities and photon emission intensities are reported in Section~\ref{sec:exct_probability}.

The $\Delta t$ distribution also contains a linear, decreasing background of random correlations. This background is most likely composed of very late after-pulses from the initial scintillation event, dark noise in the SiPMs, and predominantly the photoluminescence of the 3D printed cage. Ex-situ measurements of the photoluminescence from the cage material were performed at room temperature using a UV-VIS spectrometer and are shown in \autoref{fig:lolx-optical}. 
The photoluminescent lifetimes of the observed emission from the cage material appear to be composed of a short-lived fluorescent component which we are unable to resolve from the LXe response, and a long-lived phosphorescent component with a lifetime that was measured to be $1026 \pm 88$~ns~\cite{chana_phd_thesis}. It is likely that the low intensity of late photoluminescence from the 3D printed cage serves as the primary source of initial avalanches which induce the ExCT events.

\subsection{Systematic uncertainties} 
\label{sec:syst_unc}

\begin{table*}[t]
    \centering
    \caption{Systematic uncertainties considered, based on data and simulations. Simulation uncertainties are evaluated by comparison to a reference simulation with nominal inputs, and the combined 4~V dataset for data-driven uncertainties.
    The symbol `--' indicates negligible uncertainties, while `N/A' means not applicable.}
    \begin{tabularx}{0.9\linewidth}{l|>{\raggedright\arraybackslash}p{4cm}| >{\raggedright\arraybackslash}p{3cm}| >{\raggedright\arraybackslash}p{3cm}}
        \textbf{Name}  & \textbf{Variation} &  \textbf{Relative Uncertainty (\%) on $P_{\rightarrow/\leftarrow}$} & \textbf{Relative Uncertainty (\%) on $P_{corr,B}$} \\
        \hline
        NIR SiPM PDE &  +10\%, -30\% relative &+9, -29 (${\rightarrow/\leftarrow}$) & N/A\\
        NIR LP Transmittance & $\pm$10\% relative & $\pm$4.4 ($\rightarrow$), $\pm$0.5 ($\leftarrow$)& N/A \\
        Cage Reflectivity & Set to 50\% (-45 \% absolute)& -30 ($\rightarrow$), -40 ($\leftarrow$)& N/A \\
        SiPM Reflectivity  & Upper limit from vacuum data  (-5\% absolute)& -3 ($\rightarrow$), -- ($\leftarrow$) & N/A\\[5pt] \hline
        Pile-up Algorithm Threshold & $\pm$ 1 photon equivalent & N/A &+12, -7 \\
        Low Occupancy Window Selection & $\pm$ 200 ns & N/A & -- \\
        $\Delta t$ fit range & $\pm$ 80 ns & N/A & -- \\[5pt] \hline

        \textbf{Total Data Uncertainty} & & &\textbf{+12, -7}\\
        & & \textbf{+11, -40 ($\rightarrow$)} & \\
         \textbf{Total Simulation Uncertainty}  & & \textbf{ +9, -49 ($\leftarrow$)} & \\
    \end{tabularx}
    \label{tab:systematics}
\end{table*}

Five systematic uncertainties impacting the evaluation of $\nxe$ are considered for this analysis and are summarized in \autoref{tab:systematics}. Measurement systematics affect the  $\pcb$ value, while uncertainties in optical parameters modify $\pf$ and $\pb$ through the simulation. Uncertainties on values taken from the literature and statistical uncertainties, listed in \autoref{tab:modelinputs}, are also included as variations to inputs of the $\pcb$ model. The primary sources of simulation uncertainty come from the absence of optical property measurements, or when the conditions of existing measurements differ from the temperature or wavelength region of interest for LoLX. 

\begin{table*}[ht]
    \centering
    \caption{Input parameters for ExCT model with 1-sigma uncertainties considered in the analysis. $\drr$ is the ratio of the rates of candidate pulses between the LP and bare channels. $\pix$ is the probability of internal cross talk. $\pf$ and $\pb$ are the forward (bare~$\rightarrow$~LP) and backward (bare~$\leftarrow$~LP) transport efficiencies, respectively.}
    \label{tab:modelinputs}
    \begin{tabular}{l|c|c|c}
        \textbf{Input Parameter} & \textbf{4~V Value} & \textbf{5~V Value}  & \textbf{Source} \\
        \hline
        $\drr$ & 6.3 $\pm$ 0.9 & 7.7 $\pm$ 1.2 & Measured\\
        P$_{DiCT}$ & 5.0 $\pm$ 0.5 \% & 7.0 $\pm$ 0.7 \% & \cite{Gallina_2022} \\[3pt]
        P$_{\rightarrow}$ & 4.1 $^{+0.5}_{-1.7}$ \% & Same as 4~V & Evaluated from simulation \\[3pt]
        P$_{\leftarrow}$ & 0.6 $^{+0.1}_{-0.3}$ \% & Same as 4~V & Evaluated from simulation \\
    \end{tabular}
\end{table*}

We vary the NIR SiPM PDE by a relative +10\% to -30\% with respect to the datasheet. The larger negative variation on SiPM NIR PDE is motivated by a lower than expected efficiency observed in the Hamamatsu SiPMs for VUV LXe scintillation. In the overvoltage range in which the data were collected, the SiPM PDE is approaching saturation \cite{Gallina_2022}, so we use the nominal 4 V SiPM PDE from Hamamatsu's datasheet for all simulations. The internal PDE ($iPDE$) is calculated from this as described in Section \ref{sec:sipm-optics}. We expect that the +10\% PDE variation captures an increase in PDE at 5~V overvoltage.

The 3D printed cage material has no available reflectivity data, so the base reflectivity is computed by Geant4 using Fresnel equations based on the refractive index of the material. The specific plastic blend of the Formlabs Durable resin is proprietary but is stated to be acrylate-based \cite{formlabs}. We use the refractive index for acrylic plastic and vary the absolute reflectivity by 50\%. This systematic has the largest impact on the ExCT transport probabilities $\pf$ and $\pb$, and dominates the systematic uncertainty of the final result.

The SiPM reflectivity model described in Section \ref{sec:sipm-optics} defines the lower and upper limits for the SiPM reflectivity. In summary, the model used for the lower limit reproduces the vacuum reflectivity measured in \cite{nexo_reflectivity_vacuum}, and is used as the central value for calculation of $\pf$ and $\pb$. The more reflective case ascribes full Fresnel reflectivity to the active area of the SiPM, a strategy giving agreement with the UV reflectivity in LXe measured in \cite{Wagenpfeil_2021} (Shown in \autoref{fig:lolx-optical}. The NIR transmittance of the longpass filters is varied by $\pm 10\%$ relative to the data-sheet transmittance of 90\%. These optical systematics modify $\pf$ and $\pb$ although they are subdominant to the uncertainty on the cage reflectivity.

The LoLX data analysis depends on the low-occupancy assumption, since there must be a sufficiently small probability of observing two uncorrelated pulses within the correlation window [-24 ns, 24 ns] to resolve the percent level contribution of ExCT. One potential source of enhanced coincident light in the low-occupancy search region is pile-up from multiple source decays occurring in the same digitization window. The $^{90}$Sr source had an activity of $\sim$ 330 Bq at the time of data collection, so assuming secular equilibrium of $^{90}$Sr and $^{90}$Y, the estimated pile-up probability in the 3~$\mu$s DAQ window is $\sim$1 in 1000.

We observed significant radio frequency interference (RFI) during data-collection. RFI is present in coincidence across the majority of channels at the $\sim$SPE level. Combined with single or multiple decays, this can impact the low-occupancy region. A pile-up rejection algorithm was implemented to search for multiple trigger-candidates by inspecting all waveforms for threshold crossings. Events with more than one trigger-candidate are excluded from the analysis, which very efficiently removes RFI + physics coincidences. We vary the threshold of the pile-up algorithm by $\pm$1 PE equivalent in ADC units and measure a $\sim$ 10\% relative change to the measured $\pcb$ values. The asymmetric uncertainties from the pile-up algorithm threshold are considered in the final results as error bars on the data driven $\pcb$ value. 

The probability of direct internal cross-talk $\pix$ was taken from literature without any associated systematic. Several other effects were evaluated. This includes the previously mentioned background function fitting range, the specific location and size of the low-occupancy window, the energy of the preceding trigger event, and the temperature variations within a run. These uncertainties have a negligible impact on $\pcb$.

The fitting range of the Gaussian peak was varied by $\pm 80$ ns which resulted in a < 0.1\% change to the observed signal. This is subdominant by a factor of 10 to the other sources of systematic uncertainties listed above, and is therefore not included in the total uncertainty evaluation.

\subsection{ExCT probability and emission intensity}
\label{sec:exct_probability}

\begin{table*}[ht]
    \centering
    \caption{External cross-talk probabilities from the time correlation analysis, and the resulting mean number of photons emitted per avalanche within silicon and into LXe.}
    \label{tab:yields}
    \begin{tabular}{c|c|c|c|c}
        \textbf{Overvoltage}  & \textbf{Observable $\pcb$ (\%)} & \textbf{\small{ExCT probability ($P_{Forw}$) (\%)}} & \textbf{Emission within Si ($\gamma$/Av.) } & \textbf{Emission into LXe, $\nxe$ ($\gamma$/Av.)}\\
        \hline
        4 V  & 3.9 $\pm$ 0.2 (stat.) $^{+0.5}_{-0.3}$ (sys.) & $2.1^{+1.2}_{-1.3}$ &$20^{+11}_{-9}$  & $0.5^{+0.3}_{-0.2}$\\[5pt]
        5 V & 5.2 $\pm$ 0.2 (stat.) $^{+0.6}_{-0.4}$ (sys.) & $2.6 \pm 1.4$ &$25^{+12}_{-9}$  & $0.6^{+0.3}_{-0.2}$\\
    \end{tabular}
\end{table*}

\begin{figure}[ht]
    \centering
    \includegraphics[width=1.0\linewidth]{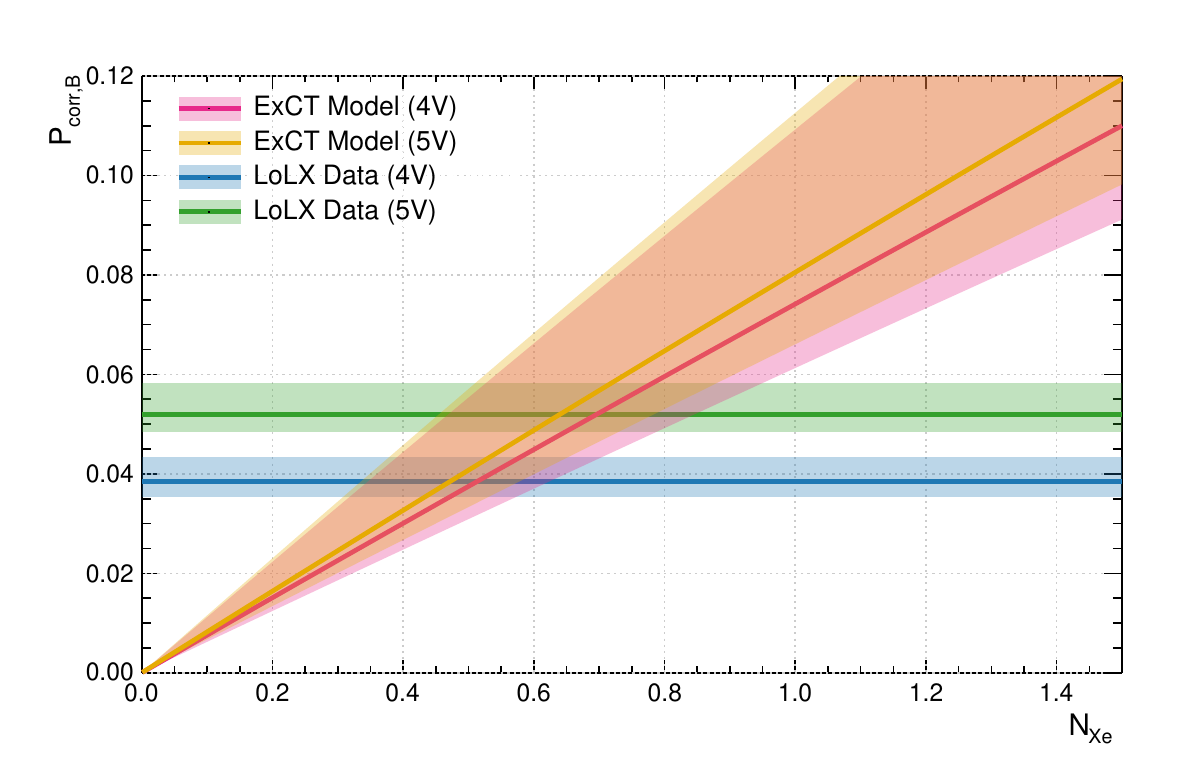}
    \caption{Effective model for correlated ExCT probability, described in Section \ref{sec:eff-model}, with measured values of correlated probability in LoLX. The mean number of photons emitted into LXe per avalanche, $N_{Xe}$ is estimated from the crossing of the data curves with the model curves for each overvoltage value. Shaded regions represent total 1-sigma uncertainty bands. LoLX data bands include statistical and systematic uncertainty in quadrature, and model uncertainties are the combined uncertainty from all sources.}
    \label{fig:ext-model}
\end{figure}

Using the experimental, simulated, and literature based inputs outlined in \autoref{tab:modelinputs}, \autoref{eq:effective-model} and \autoref{eq:rate-ratio} are solved numerically to find the $\nxe$ value that agrees with the experimentally measured $\pcb$. The result is shown in \autoref{fig:ext-model}, where $\nxe$ is given by the intersection of the model curves and the horizontal data lines. This yields the average number of photons emitted into LXe per avalanche $\nxe$ as 0.51 and 0.64, for the 4~V and 5~V data, respectively. Values with uncertainties are reported in \autoref{tab:yields}.

The shaded bands in \autoref{fig:ext-model} represent the total uncertainty on the model or data. All systematic uncertainties impacting $\pf, \pb$ are added in quadrature, yielding the upper and lower bounds given in \autoref{tab:systematics}. The error on $\drr$, the data driven rate ratio, was determined from the statistical variance across runs. The error on the internal cross-talk probability $\pix$ is taken from literature. The error on $\pcb$ comes from the statistical uncertainty and the asymmetric pile-up algorithm systematic. When solving for $\nxe$ the uncertainty is determined by adding in quadrature the difference from evaluating the model at the upper and lower data values (intercepts between model and data), with the difference calculated from evaluating the central data values with the upper and lower models defining the shaded regions in \autoref{fig:ext-model}.

The simulated transport and detection probabilities $\pf$ and $\pb$ preserve the correct angular and wavelength characteristics of the ExCT process in LoLX due to the simulation strategy, which maintains consistent wavelength and angular emission independent of photon intensity. Therefore within LoLX, the true directional ExCT probability for either the forward or backward process is given by $P_{Forw}$ or $P_{Back}$. The probability for the forward process, a bare SiPM triggering any LP SiPM, is roughly $P_{Bare \rightarrow LP} \approx \nxe \pf \approx 0.5 \times 0.04 \approx 2\%$. To reiterate, $\pf$ is detector dependent, owing to the highly non-isotropic emission from the SiPMs, and optical filtering and detector geometry of LoLX.

\subsection{ExCT internal emission intensity}

To calculate the ExCT intensity for SiPMs in a different medium or using an optical model different than \autoref{fig:ext-emitted}, the photon yield inside the silicon $N_{Si}$ is estimated. 
The fraction of photons that escape from Si and enter LXe is considerably small due to the large refractive index mismatch between Si and the SiO$_{2}$ surface layer ($\sim$3.78 vs $\sim$1.46 for Si vs SiO$_{2}$ at 700 nm), resulting in the majority of photons undergoing total internal reflection. This fraction is $94.7$\% for the outgoing hemisphere, or 97.4 \% in 4$\pi$ which is evaluated by averaging the ExCT wavelength emission spectrum over the optical transmission model, for isotropic emission in silicon, as described in Section \ref{sec:ext-sim-procedure}. Dividing $\nxe$ by this average emission probability yields the number of photons produced in silicon, which is reported in~\autoref{tab:yields}. This detector-independent value can be used for modelling ExCT emission for SiPMs operated in LXe or other media.

\section{Discussion and conclusion}
\label{sec:4}
We have measured ExCT of SiPMs in a LXe detector under nominal operating conditions. This was carried out at two different values of SiPM overvoltage, with an enhanced probability observed at higher overvoltage. The results are given in~\autoref{tab:yields}.  $\pcb$ represents the probability that given a single SPAD avalanche on any bare SiPM, we observe a correlated single-PE pulse on any LP filtered SiPM. $P_{Forw}$ is the ExCT probability for the forward, bare-to-LP process, where the contribution from the `backwards' LP-to-bare process has been removed from $\pcb$ using the method described in Section \ref{sec:eff-model}. In other words, $P_{Forw}$ represents the probability that for an SPE avalanche in a bare SiPM, a single ExCT photon is detected in any LP channel in LoLX. A similar result is expected to apply to any LXe detector using the same SiPMs, with similar photocoverage or angular acceptance. Thus, we determined that for SiPMs in a high photocoverage detector, ExCT is approximately a percent-level process.

To extract a detector-independent yield $\nxe$, and to compare to ex-situ measurements of ExCT emission from SiPMs~\cite{raymond2024stimulated}, we developed a custom Geant4 physics process to simulate the ExCT emission and transport within LoLX. This simulation is available for other experiments through a public GitHub package \cite{dgallacher1:g4-ext}. This simulation also includes our more detailed SiPM response and surface optics, accounting for the differences between vacuum measurements of SiPM performance and the expected behaviour in LXe. The simulation was used to evaluate the combined transport and detection efficiency for ExCT light within LoLX and to evaluate the impact of simulated optical uncertainties. A mathematical framework was built to correct for the contributions from the different ExCT processes. This produced detector independent photon yields in LXe from a single SPAD avalanche at 4~V and 5~V overvoltage. The emission intensity of ExCT is expected to follow the gain of the SiPM which is proportional to the overvoltage as observed in this study. The ratio of $\nxe$ at 5V/4V is $\sim$1.25, identical to the ratio expected from overvoltage dependence. 

The emission intensities given in \autoref{tab:yields} are for photons created within the silicon of the device. The internal yield values reported here are in tension with the reported intensities of the Hamamatsu VUV4 SiPMs from ex-situ measurements of SiPM ExCT of 48.5$\pm$9.5 \cite{raymond2024stimulated} at 4~V overvoltage. After correcting for the optics in LXe, the estimated number of photons emitted into LXe from \cite{raymond2024stimulated} is 1.55$\pm$0.62, compared to 0.51$^{+0.29}_{-0.24}$ measured at 4~V in this study, a difference of factor $\sim$3. The angular acceptance of the measurement in \cite{raymond2024stimulated} is $\sim 27\degree$ (NA = 0.45) above the SiPM, while in the LoLX detector we are sensitive to the full emission angle distribution averaged over the geometry of the detector. The optics corrections to convert from vacuum or LXe to an internal photon yield make the simplifying assumption of treating the SiPM as a flat surface, and ignoring any possible photon absorption on exit. This also ignores possible effects of the SiPM's microstructure and individual SPAD 2D geometry on photon emission. This may contribute to an overestimate of the large-angle emission in extrapolating the ex-situ measurements to 2$\pi$, which could contribute to this discrepancy. Another source of inaccuracy may be the assumption of isotropic ExCT photon production in the silicon, as some fraction of the ExCT photons are likely produced via bremsstrahlung radiation, which is directional at higher energies. A further source of discrepancy may be that comparing vacuum and LXe measurements involves transforming between the photon intensities outside and internal to the silicon, in order to factor out the transmission differences due to the different media. Boundary effects, and photon absorption on the SiPM microstructure may impact the photon production and escape mechanism in a manner not captured by this simplified model of factoring out the optics, as LoLX is sensitive to a much broader angular acceptance than the microscope based measurement of \cite{raymond2024stimulated}.

Follow-up measurements to constrain the angular emission of ExCT photons from SiPMs are required to resolve this tension. A more involved analysis using specific SiPM locations and geometric correlations may help constrain the angular behaviour of ExCT in LoLX. In addition, improved understanding of the interplay between the `forward' and `backward' processes will inform future studies. In analyzing \autoref{eq:effective-model} it can be deduced that the rate ratio $R_R$ drives the backwards process' contribution to the observable $\pcb$. Due to the high pulse rate in the LP channels, the backwards process has a sizeable contribution to the signal, which was not initially expected. For future studies, other channel combinations or normalization procedures are likely to be more straightforward or effective in probing the ExCT process.

The primary sources of systematic uncertainty in this analysis originate from the absence of measurements of optical properties at operating conditions for LXe. The extrapolation from room temperature and air or vacuum conditions to 175 K and LXe is required for most measured optical parameters. This motivates the need for cryogenic optical characterization and testing facilities. For our application and similar large physics experiments, this requires both vacuum cryogenic measurements and testing in LXe.

Since this data taking campaign, LoLX has undergone major upgrades where the 3D-printed fluorescing detector structure has been replaced by a simplified cubic geometry of SiPMs mounted on circuit boards, as well as a complete overhaul to the cooling system to enable longer duration runs. RFI pickup was later corrected with more thorough grounding and avoidance of ground-loops in the front-end electronics system.

Future measurements will include angular correlation evaluations and a comparison to the simulated optical emission shown in \autoref{fig:ext-emitted}. The upgraded LoLX detector also includes an upgraded DAQ, on-loan from the MEG-II collaboration,  with up to 5 GHz digitization rates \cite{GALLI2019399} which are expected to improve on the signal-to-noise for a follow-up ExCT measurement. We plan to perform improved measurements, enabling individual channel correlations to sample specific angular acceptances. By measuring the ExCT correlation probability for different viewing angles between SiPMs we aim to characterize the angular dependencies of the light emission and compare to the simulation models. 

The impact of ExCT for future large-scale rare-event search experiments will be evaluated in future work informed by the effective model and simulation framework developed here. This work solidifies the expectation that ExCT is a non-negligible, percent level process depending on detector optics, photocoverage, and operating voltage. Future work will focus on evaluating the effect on energy resolution, low energy trigger thresholds, and analyses requiring timing correlations. Understanding the effect of SiPM ExCT on detector performance variables is critical for their use in future rare event search experiments.

\begin{acknowledgements}
This research was undertaken thanks in part to funding from the Canada First Research Excellence Fund through the Arthur B. McDonald Astroparticle Physics Research Institute, with support from the Natural Sciences and Engineering Research Council of Canada (NSERC), the Fonds de Recherche du Qu\'ebec (FRQ), and the Canada Foundation for Innovation (CFI).
\end{acknowledgements}



\bibliographystyle{spphys}       
\bibliography{refs}   

\end{document}